\documentstyle[aps,manuscript,psfig]{revtex}
\begin{document}

\title{\bf  Hysteresis at low Reynolds number: the onset of 2D vortex
shedding}
\author{V.K.~Horv\'ath, J.R.~Cressman, W.I.~Goldburg, and
X.L.~Wu}
\address{Department of Physics and Astronomy, University of
Pittsburgh, Pittsburgh, PA 15260}


\maketitle

\bigskip

\begin{abstract}
Hysteresis has been observed in a study of the transition between
laminar flow and vortex shedding in a quasi-two dimensional
system. The system is a vertical, rapidly flowing soap film which
is penetrated by a rod oriented perpendicular to the film plane.
Our experiments show that the transition 
from laminar flow
to a periodic K\'arm\'an vortex street can be hysteretic, i.e.
vortices can survive at velocities lower than the velocity needed to
generate them.
\end{abstract}

Pattern forming fluid flow is often difficult to understand
even at quite modest velocities. A widely studied example is the
chain of vortices generated downstream from a rod placed in a
uniform flow, with the axis of the rod oriented perpendicular to
the flow direction. In a pioneering study of this problem,
K\'arm\'an analyzed the stability of this vortex
street\cite{karman}.  His starting point was the assumption that the flow
is two dimensional, i.e., no velocity variations are generated
along the rod axis. He was concerned only with the stability of the
vortex array and made no attempt to explore the transition itself.
Although researchers have put considerable effort to better
understand this phenomenon, which is
considered to be a precursor to turbulence,
a full understanding is still lacking\cite{comment:2}.

It is now widely accepted that the transition (primary instability)
from laminar flow to vortex shedding is supercritical in nature; on
changing the mean flow speed $\overline{V}$ around the transition
point, vortex shedding appears or disappears at a unique
velocity $V_{c}$. 
Below (above) $V_{c}$ the only stable state
of the system is that of the laminar (vortex shedding) flow.

In an important wind tunnel experiment on vortex shedding, Provansal
et al.~\cite{provansal:87} showed that the primary instability
occurs at a critical Reynolds number $Re_{c}\simeq$46.
The vortex shedding disappeared at the same velocity at which 
the onset occurred, i.e. hysteresis was absent.
These measurements are well described by the Landau-Stuart theory
for supercritical transitions.
Sreenivasan et al.\cite{ghia}  observed that an acoustic
disturbance or a vibration of the rod itself could activate the vortex
street when $Re$ was less than $Re_{c}$ but greater than some minimum
value $Re_{min}\simeq Re_{c}/2$.  
Their experiment showed that the general behavior
of the system was position independent.  
The Landau-Stuart theory was adequate to explain their 
experimental findings too.
Couder\cite{Couder:86} and Gharib\cite{gharib:89} were among the first
to study vortex shedding behind a cylinder in soap films.  In
particular, Gharib and his associates carried out systematic
investigations of the vortex shedding frequency $f$ as a function of
$\overline{V}$ in a horizontally flowing film\cite{gharib:89}.  

In this paper we present measurements demonstrating that the
primary instability can be hysteretic.
This finding is also in contrast with numerical
results\cite{henderson:96}.
Although hysteretic transitions are well known in
fluid dynamics\cite{brika:93}, we know of no published
laboratory experiments or computer simulations, which suggest that
the onset of vortex shedding from a rigid rod
is a hysteretic phenomenon.

Our experimental setup consists of a
rapidly flowing soap film formed between
two vertically positioned 0.25 mm diameter nylon lines
separated by 6 cm (see Fig.~1).
The lines are parallel along a 45 cm vertical segment.
They are joined at both ends of the segment.
The film is fed continuously from the top
through a high precision metering valve.
The valve was opened and closed at different rates by a motor
connected to it, thus the acceleration
$d\overline{V}/dt$ was under experimental control.
The film is a mixture of soap (Dawn) and distilled water
with a typical soap concentration of 1 wt.\%.

Using an infrared absorption technique, the soap film thickness profiles
parallel and perpendicular to $\overline{V}$ were measured. 
In the absence of
the rod, the horizontal film thickness was essentially flat with less
than 5\% variations, whereas there was systematic thinning along the
vertical direction. 
Due to a small acceleration of the film,
the latter amounts to a 1\% change in the thickness
over a distance between the two LDV probes. 
The average thickness, which was found to be several microns,
does not change over the range of
$\overline{V}$ spanned by our experiments.

The velocity measurements were made using a dual head
laser Doppler velocimeter (LDV).
At the center of the parallel
segment of the soap film, a glass rod of diameter $d$=1 mm penetrates the film
in the $z$ direction.
The velocities $\overline{V}$
and the fluctuating  horizontal component $V_{x}(t)$
were measured at points 5 cm above and 5 cm
below the rod respectively.

At a certain critical flow rate there is a
transition from the laminar flow (LF) state to 
the vortex shedding (VS) state.
Slightly below this critical 
velocity, the
flow is characterized by two counter-rotating vortices
underneath the rod (see Fig.~2a). 
In the VS state 
these recirculating vortices peel off the rod and flow
downstream. The continuously generated counter-rotating vortices
form a vortex street (see Fig.~2b), as described by 
K\'arm\'an\cite{karman}.

All measurements were made near the
transition.  The experiments start with a value of
$\overline{V}$ corresponding to the LF regime.  
The valve is then
opened to slowly increase $\overline{V}$.
The critical velocity where VS
commences is called $V_{c}^{up}$.  
After 
some waiting time 
the valve is slowly closed.
As a result $\overline{V}(t)$ decreases, and at
$\overline{V}(t)$=$V_{c}^{down}$ the system undergoes a transition
from VS into LF. 
This cycle was repeated several times in each run.
To find the transition from LF to VS, $V_x(t)$ was recorded 
together with $\overline{V}$.
In theory $V_x(t)$ is constant in the laminar state and equal to
zero directly below the center of the rod.  
In the VS regime $V_x(t)$ is
expected to display periodic behavior.

In order to determine the transition velocity accurately we have
calculated the velocity probability distribution function $P(V_x)$.  
For this purpose the total recording time was
segmented into intervals $\Delta t\simeq$200\/ms.  
Then $P(V_x)$ was calculated
within each interval.  The intervals $\Delta t$ were much smaller
than the characteristic changing time of
$\overline{V}(t)$
but large enough to include a statistically
significant number of data points
and many periods of oscillation in the VS state. 
A typical result is shown in Fig.~3. 
Here the magnitude of $P(V_x,t)$ is mapped into gray scale values
and shown as a function of $V_x$ and $t$.
It can be seen
that sharp changes in $P(V_x,t)$ precisely indicate the transition
between the LF and VS regimes.  
In the LF regime, $P$ is sharply peaked
at $V_{x}$=0, but in the VS regime $P$ has a broad distribution,
as for a (noisy) sine wave.
These measurements provide a
powerful way to accurately determine the times $t_d^*$ and $t_u^*$ of  
LF$\to$VS and VS$\to$LF
transitions\cite{comment:1}.

The hysteresis is apparent in
Fig.~3, where $\overline{V}(t)$ and $P(V_x,t)$ are displayed
together.
By determining the value of $\overline{V}(t)$ at the
transition times $t_u^*$ and $t_d^*$ one find that $V_c^{up}=\overline{V}(t_u^*)$ 
is not equal to $V_c^{down}=\overline{V}(t_d^*)$,
that is, there is no unique critical velocity for the transition.
In this analysis we have incorporated the fact that 
$\overline{V}$ is noisy at the level of $\pm 4\%$.
Therefore to determine $V_c^{down}$ (and $V_c^{up}$) we always used 
the lowest (and highest) 
values of (the fluctuating) ${\overline V}$ at the transition.

The critical velocities were measured in 20 experimental runs.
The ensemble average of the 
different experimental runs provide
$<V_c^{down}>=0.42\pm0.02$\/m/s.
The relative difference
$(<V_c^{up}>$$-$$<V_c^{down}>)/<V_c^{down}>$ is approximately 14\%.
The Reynolds number -which is defined as $Re=Vd/\nu$-
is roughly 50 in the hysteretic gap.
A large uncertainty of this value comes from the 
poorly determined two-dimensional 
soap film viscosity $\nu$, which depends on the 
film thickness\cite{Rivera_forever}.

Next, we demonstrate that the observed hysteresis is {\it not} a result
of some delay in the response of the system to the bifurcation,
as it would be the case for dynamical hysteresis\cite{strogatz:97,berglund:97}.
We set the mean flow rate constant in the interval 
$[V_c^{down}$,$V_c^{up}]$.
In this case the laminar state can persist for an indefinite length of
time.
However applying 
sufficiently large acoustic or pulse-like mechanical perturbations, 
the system could be driven into the VS state. 
The system remains in this state even if the perturbation is turned off.
Applying similar perturbations VS$\to$LF transition was never observed
at constant $\overline{V}$.
As a result we can conclude that we observed static hysteresis,
i.e. its existence is independent of the rate of change of
$\overline{V}$.

At the transition, a certain time interval is required for the
new steady state to establish itself as it is seen in the 
close vicinity of $t_u^*$ in Fig.~3.
The overall time of the transient remains finite even
if $\overline{V}$ is changed very slowly.
This would not be true for continuous transitions\cite{strogatz}.
Careful examination of $P(V_x,t)$ shows
that the most probable $V_x$ grows exponentially at the
transition points.
Using exponential curve fitting one
finds that ${\tau}^{growth}\simeq$ 1.32$\pm$ 0.05 s.
Thus it was necessary to vary $\overline{V}$
slowly enough, that this finite transit time  did not interfere
with our identification of $V_c^{up}$ and $V_c^{down}$.

Because hysteresis  is not seen at the onset of vortex shedding
in 3D systems, we have looked for possible side effects
that might be responsible for our unexpected finding.
It is known that oscillations
of the rod can induce hysteresis \cite{brika:93}.
Our rod is rigidly  supported,  so  its  vibration  can  only  be 
induced by the fluid itself \cite{rivera}.
Since the film is only a few microns thick, it carries insufficient
momentum flux
to  deflect  our  1  mm  diameter glass rod more than a nanometer.
The  possible  motion  of the vertical fishing lines which
support the film  was  also  considered.
Taking into account the geometry of our experimental setup,
we  estimate  that their deflection,
produced by the passage of an eddy, is no more than a micron.
Auxiliary experiments were also performed to exclude wetting properties
and the effect 
of the air boundary near the film 
as possible sources of 
hysteresis \cite{longversion}.

Our experiments are the first to establish that the transition to
vortex shedding can be subcritical. 
In the absence of any
available theory, we consider  a generic model based on
an amplitude equation\cite{landau,stuart},
namely a fifth order Landau
equation\cite{strogatz} for the complex velocity
$\tilde v$=$v_{x}+iv_{y}$:
\begin{equation}
   d{\tilde v}(t)/dt={\tilde{a}\tilde v}
   +2{\tilde{b}}v^2{\tilde v}-{\tilde c}v^4{\tilde v} ,
   \label{eqcomplex}
\end{equation}
where all variables indicated with tilde are complex
and $v_i=V_i-<V_i>_t$ ($i$=$x$ or $y$).
The complex velocity 
can be written as ${\tilde v}$=$ve^{i\phi}$, where
$\phi$ is the phase factor and $v$ is the amplitude.
In this case the evolution of $v$ is governed by the real part of
Eq.~(\ref{eqcomplex}):
\begin{equation}
   dv/dt=v(a_{r}+2b_{r}v^2-c_{r}v^{4}) ,
   \label{eqreal}
\end{equation}
where the subscript $r$ is used to designate the  real
part of complex numbers and $a_{r}$ is considered to
be proportional to $(\overline{V}-V_{c}^{up})$.
In the experiments $a_{r}$ is adjusted by varying
$\overline{V}$\cite{comment:4}.

In the following we consider the stationary solutions of
Eq.~(\ref{eqreal}). 
The laminar solution is $v_{LF}=0$.
Note that at the position where $V_x$ is measured,
$\langle$$V_x$$\rangle_t=$$0$.
Therefore $v_x(t)=V_x(t)$ for that particular point.
The time-dependent solution
is described by: 
\begin{equation}
   {v_{VS}^{\pm}}^2=
  {b_{r}\over c_{r}} ({1\pm\sqrt{1+a_r\/c_r/b_r^2}}).
  \label{v_s}
\end{equation}
It suffices to retain only the positive values
of Eq.(\ref{v_s}).
Since the minus sign in this equation corresponds 
to an unstable state,
the observable  vortex shedding is represented by $v_{VS}^{+}$.

After some elementary algebra one can draw the stability (or
bifurcation) diagram of this model (see Fig.~4, where only the 
positive solutions are shown).
The solution $v_{LF}$ is stable only if ${a_r}$ is smaller than
a critical value $(a_r)_c^{up}$=0
(see regimes I and II).
The solutions $v_{VS}^{\pm}$ exist only 
for ${a_r}${$\ge$}$(a_r)_c^{down}$, where $(a_r)_c^{down}$=$-b_r^{2}/c_r$.

If $a_r<(a_r)_c^{down}$ then the system is in the LF state.
On increasing $a_r$, the system enters into the 
hysteretic regime (II), where the system 
can stay in the LF state as long as the amplitude of the
fluctuations does not exceed the value of $v_{VS}^{-}$.
If $a_r>(a_r)_c^{up}$ then the LF state becomes unstable;
even infinitesimally small noise will force the system
to switch from $v_{LF}$ to $v_{VS}^{+}$.
In regime II, finite perturbations with amplitude larger than
$v_{VS}^-$ force the system to switch to the VS state.
Such a switch is clearly inconsistent 
with a supercritical bifurcation, for which the
saturation amplitude
continuously changes with the control parameter.
The everywhere stable solution $v_{VS}^{+}$ does not exist below
$(a_r)_c^{down}$.
Therefore the system must return from the VS state 
to the LF state   at
$a_r=(a_r)_c^{down}$ even when noise is absent.
In this sense the VS$\to$LF transition is more robust
than its inverse.

Our observations are consistent with all the above described
features of the model. However the application of this model 
for data fitting purposes is not obvious, since the relationship between
the parameters $a_r,b_r,c_r$ and the control parameter
of the experiment is unknown.

We use
the above model and choose $a_r$ 
to be $a_r=a_1\/\epsilon$, where $\epsilon=(\overline{V}-V_c^{up})/V_c^{up}$. 
Other parameters are kept
as fitting constants. 
In this case the relevant fitting parameters are
$a_1/c_r$, $b_r/c_r$, and $V_c^{up}$.
Figure~5 shows the results of our fits together with the 
same experimental data set presented in Fig.~3.
Shown here is the $\overline{V}$-dependence of the values of $v_x$
for which $P(v_x)$ is maximum.
This figure demonstrates both the sharpness of
the VS$\to$LF transition and the breadth of the
LF$\to$VS transition in our experiment.
Open symbols show the measurements with decreasing $\overline V$
and closed symbols refer to  measurements made in the
opposite direction.

The dotted line in Figure 5. is the best fit to the data
with $V_c$ fixed at a value of 0.504 m/s, which is
the largest mean velocity where
the laminar state was still observed.
This fit seriously underestimates the width
of the observed hysteretic gap.
The dashed line is the best fit with a
different constraint:
it was required
that the fit must terminate exactly at the
lowest velocity for which the VS state exists. 
In this case the part of the curve which represents
the unstable trajectory, terminates 
on the laminar branch in the middle of the hysteresis loop, 
predicting that LF state should not exist
above $\overline V$=$0.45m/s$.
The solid line designates a fit where
both of the above constraints are imposed.

Although the generic model of the hysteresis 
is in fair agreement with our observations,
none of the above fits are very satisfactory.
It is interesting to note that 
a simple three-parameter 
phenomenological equation 
$P_0({\overline V}-V_c)+2P_1v-P_2v^{2}=0$ 
provides a surprisingly good fit to our data in the VS state. 
Using least square fitting to the solution 
of this equation provides the following result:
$v_x=0.04\pm\sqrt{\overline{V}/39.06-0.0113} $,
where the velocities are in units of $m/s$.

In summary, we have observed hysteresis at the onset of the 
K\'arm\'an vortex street in a 
quasi two-dimensional film.
In view of prior studies
of vortex
shedding 
this result is surprising.
Although the origin of the hysteresis is still not clear,
we can exclude mechanical vibrations, wetting properties of the rod, and
the air boundary layer as possible sources.
Our preliminary experiments suggest
that the effect may be connected with
the fact that (a) the film is very thin in the 
recirculation region just before vortex shedding starts
and (b) the kinematic viscosity of the film
depends on its thickness\cite{Rivera_forever,Rutgers,Ecke}.
The fifth order amplitude equation is in qualitative
agreement with our observations, but a simple phenomenological
equation provides a better numerical fit to the experimental data
in the VS regime.
Theoretical explanation of our experimental findings
as well as the full 
description of the vortex shedding continues to offer
a great challenge for the future.

We thank M.~Rivera for his comments and help in the experiments.
We have profited from discussions and kind correspondence
with M.~Gharib, R.~D.~Henderson, D.~M.~Jasnow, and K.~R.~Sreenivasan.
This material is based upon work supported by the 
North Atlantic Treaty Organization
under a Grant DGE-9804461 awarded to W.I.~Goldburg and V.K.~Horv\'ath. 
The work was also supported by NSF grant DMR-9622699, 
NASA grant 96-HEDS-01-098 and by additional support
from the Hungarian Science Foundation grant OTKA F17310.


\begin{references}
\bibitem{comment:1}
In identifying the transition time it must be taken into account that
  $\overline{V}(t)$ and $v_{x}(t)$ is measured different places. \ Since
  $\overline{V}(t)$ near the transition is $\sim$50 cm/s, the related time
  delay (about 0.2 s) is negligible to the characteristic changing time of
  $\overline{V}$.
\bibitem{comment:4}
The mean velocity {$\overline{V}$} is used here rather than the Reynolds number
  because of an uncertainty concerning the value of the viscosity of the
  flowing soap film.
\end{references}


\newpage
\centerline{\Large\bf Figure captions}
\begin{enumerate}

\item{The schematic of the experimental setup.
     The diameter of the rod is 1\/mm.
}

\item{Interference pictures of the flowing soap film below the
     rod in (a) the laminar state and (b) the vortex shedding state.
     It was shown previously
     that the thickness of the
     film is correlated with the film velocity [10].
}

\item{Here the function $\overline{V}(t)$ is shown together
     with the $P(v_x,t)$ map at
     $|d\overline{V}/dt|$=0.004 m/s$^2$.
     The gray scale corresponds to the probability density of $v_x$.
     The solid and dashed lines are intended to guide the eye for the
     transitions VS$\to$LF and LF$\to$VS.
}

\item{The bifurcation diagram of Eq.~(\ref{eqreal}).
     The turning point of the
     nontrivial solution is indicated by {\bf A}. 
     The solid and dashed lines designate the
     stable and unstable solutions respectively.
}

\item{The main figure shows our  
 experimental data together with 3 different fits
 using Eq. (\ref{v_s}).
 The circles and squares represent two different data sets.
 Closed and opened symbols are used for data taken at increasing and
 decreasing $\overline{V}$ respectively. 
 The solid and dashed lines have the same meaning as in Figure 4.
 The inset shows our best parametric fit to the data (see text).
}

\end{enumerate}
\newpage

\renewcommand{\thepage}{}
\renewcommand{\figure}{}

\begin{figure}
   \psfull
   \centerline{\psfig{figure=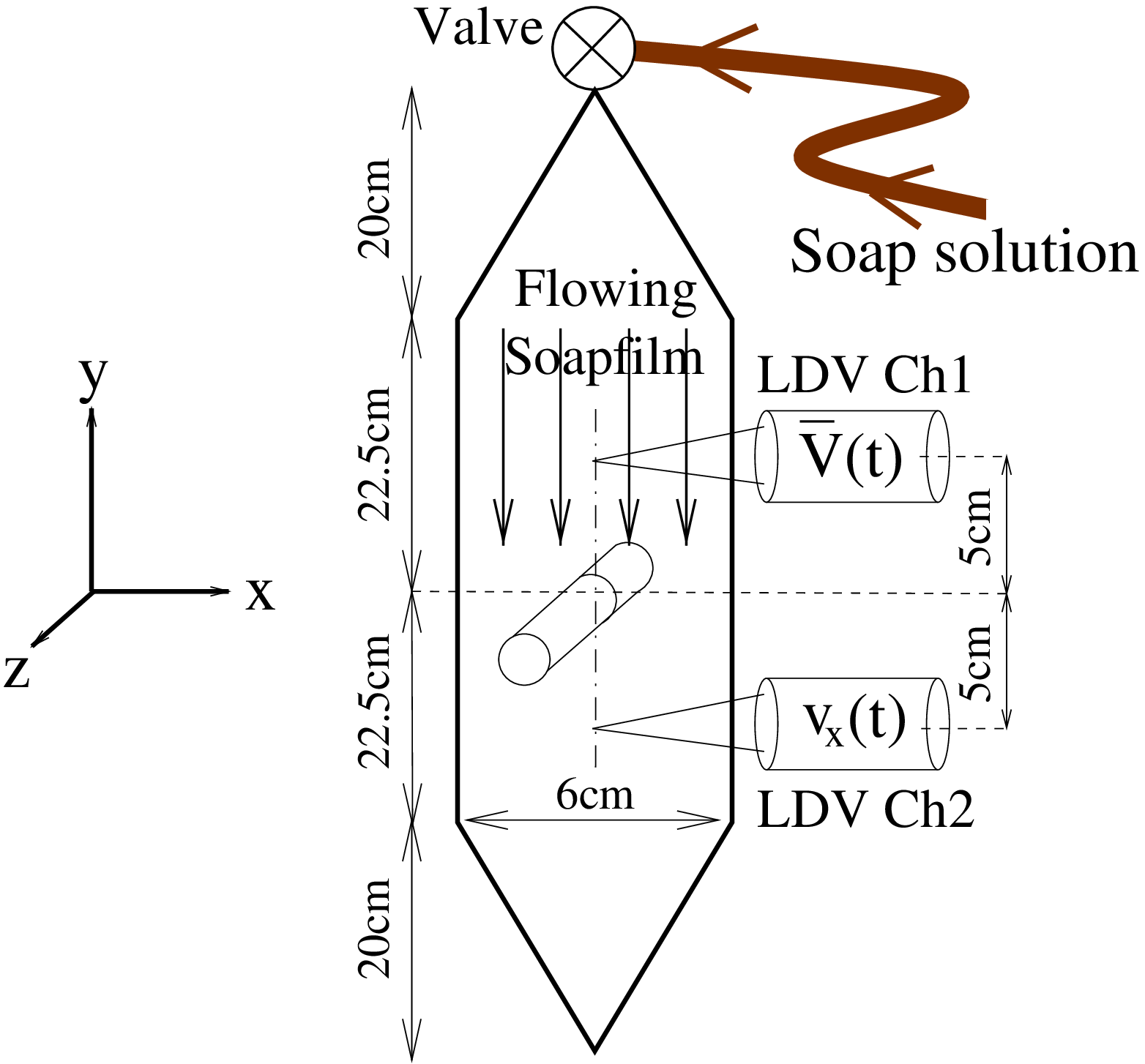,height=13truecm,clip=true}}
\end{figure}
\vspace*{\fill}
\begin{flushright}
{\LARGE\bf Figure 1}
\end{flushright}
\newpage

\begin{figure}
   \psfull
   \centerline{\psfig{figure=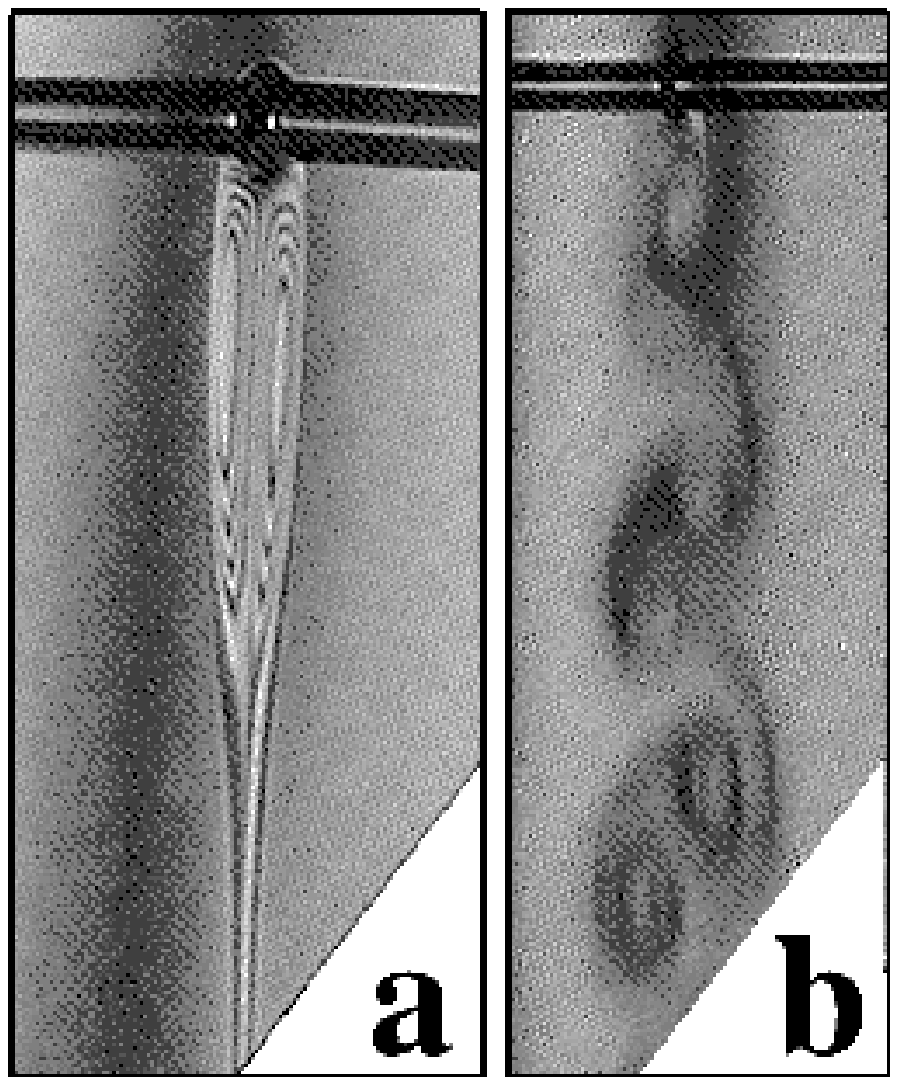,height=18truecm,clip=true}}
\end{figure}
\vspace*{\fill}
\begin{flushright}
{\LARGE\bf Figure 2}
\end{flushright}
\newpage

\begin{figure}
   \psfull
   \centerline{\psfig{figure=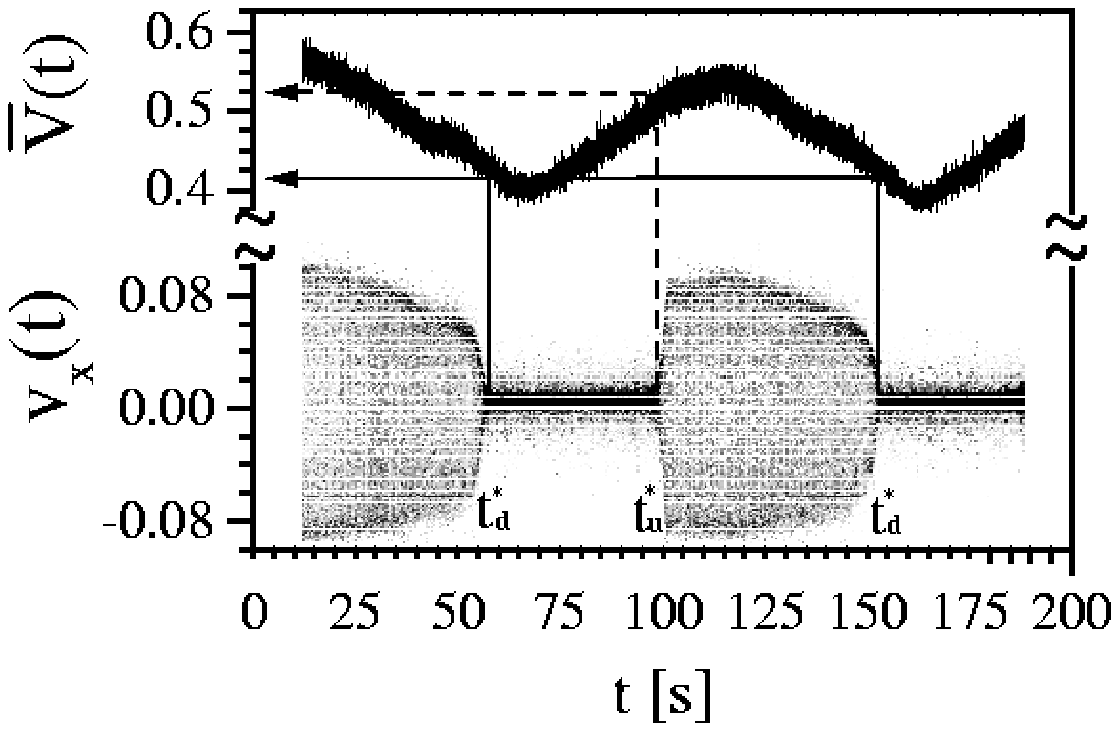,width=14truecm,clip=true}}
\end{figure}
\vspace*{\fill}
\begin{flushright}
{\LARGE\bf Figure 3}
\end{flushright}
\newpage

\begin{figure}
   \psfull
   \centerline{\psfig{figure=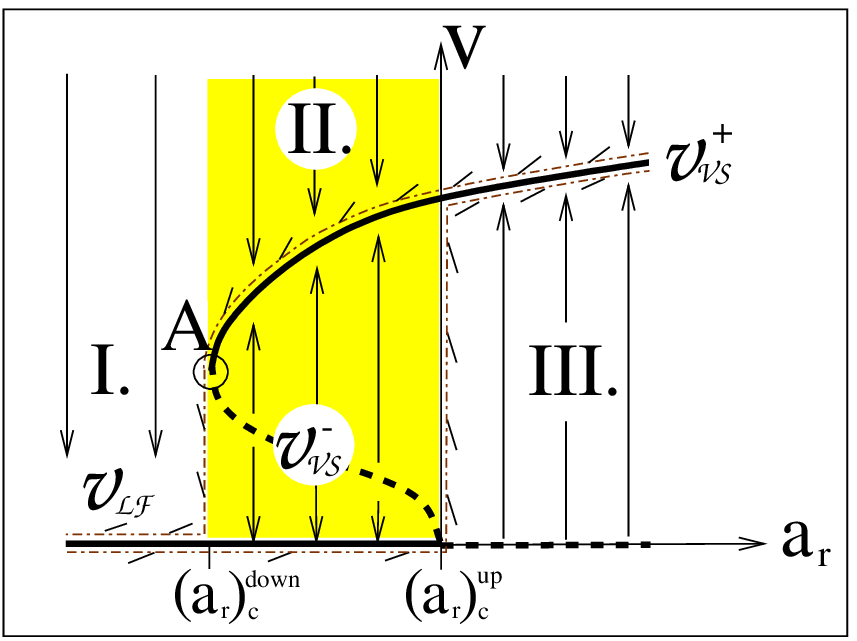,width=14truecm,clip=true}}
\end{figure}
\vspace*{\fill}
\begin{flushright}
{\LARGE\bf Figure 4}
\end{flushright}
\newpage

\begin{figure}
   \psfull
   \centerline{\psfig{figure=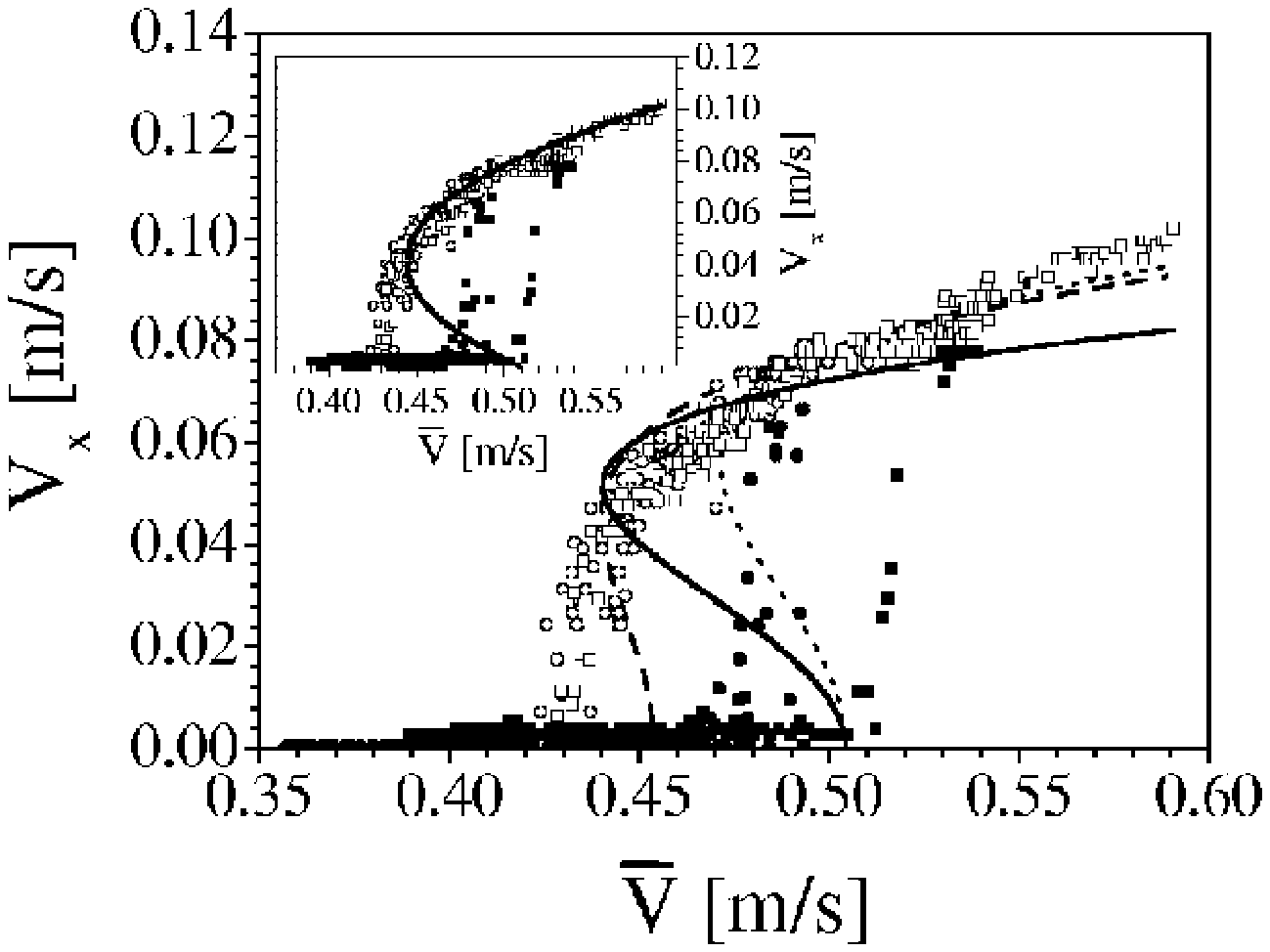,width=14truecm,clip=true}}
\end{figure}
\vspace*{\fill}
\begin{flushright}
{\LARGE\bf Figure 5}
\end{flushright}
\newpage

\end{document}